\documentstyle[aps,epsf,prl,multicol]{revtex}
\draft

\begin{document}

\title{Fluctuations in a fluid with a thermal gradient} 

\author{Jaechul Oh and Guenter Ahlers}

\address{Department of Physics and iQUEST, University of
California, Santa Barbara, CA  93106, USA}

\date{\today}\maketitle

\begin{abstract}
We report measurements of the temperature fluctuations in a horizontal layer of a pure fluid confined between parallel plates and heated from below. Consistent with earlier work we found that the structure factor $S_T(q)$ (the square of the modulus of the Fourier transform of the temperature field) was consistent with $S_T(q) \sim q^{-4}$ for large horizontal wavenumbers $q$. As $q$ decreased, $S(q)$ increased less rapidly, passed through a maximum, and then approached zero. The results agree qualitatively, but not quantitatively, with the predictions of Ortiz de Z\'arate and Sengers. 
\end{abstract}


\begin{multicols}{2}

The enhancement of fluctuations in a fluid due to non-equilibrium conditions has long been an important topic in statistical mechanics. Even for the more specific case of a fluid with a thermal gradient this problem has a considerable history. \cite{FN,OS01,OS02} For large wavenumbers $q$ perpendicular to the gradient it was shown two decades ago that the structure factor $S(q)$ (proportional to the square of the modulus of the Fourier 
transform of density fluctuations) should vary as $q^{-4}$. \cite{KCD82,RP82,SC85,Sc88,LS89} This was confirmed experimentally by small-angle light scattering. \cite{LSGS90,SGSL92,LSGS94,LZSGO98} It was soon realized theoretically that the small-$q$ divergence saturates at a constant value as $q$ vanishes because of gravity. \cite{SSS93,SS93} This saturation effect was indeed observed by light scattering at very small angles from binary mixtures where concentration fluctuations are expected to play a role similar to temperature fluctuations in a pure fluid and give a much stronger
 signal. \cite{VG96,VG97} For a sample confined between parallel plates an additional saturation effect is expected to come into play because fluctuations orthogonal to the plates are cut off by the finite plate spacing $d$. \cite{OPS01,OM01,OS01,OS02}. With decreasing $q$ this latter effect is expected to lead to a maximum of $S(q)$ , with a subsequent small-$q$ regime where $S(q) \sim q^2$. However, to our knowledge this maximum and the decrease at even smaller $q$ was not accessible to prior 
experiments. Very recently Ortiz de Z\'arate and Sengers (OZS) \cite{OS02} calculated quantitatively the shape of $S(q)$ over the entire range of $q$ for realistic no-slip top and bottom boundary conditions, using the linearized Boussinesq equations of fluid mechanics \cite{Ch61} and a low-order Galerkin approximation for the fluctuations in the vertical direction. 

Here we report measurements of $S_T(q)$ of temperature fluctuations which are related to $S(q)$ by a thermodynamic factor. We covered the range $1 \alt q \alt 6.5$ where $q = 2 \pi / \lambda$ with $\lambda$ the horizontal wavelength of the fluctuations in units of $d$. We used a shadowgraph method and a thin horizontal layer of fluid heated from below. The sample was compressed SF$_6$ on the critical isochore above the critical temperature $T_c = 45.567^\circ$C where fluctuation effects become exceptionally large, and where the shadowgraph method becomes very sensitive because the refractive index $n$ has a large temperature dependence $(\partial n/\partial T)_P$. We found that $S_T(q)$ had the expected maximum in the range $3 \alt q \alt 4$, and that it decreased for larger $q$ in a way consistent with $S_T(q) \sim q^{-4}$ as predicted and previously observed. Below the maximum $S_T(q)$ tended toward zero with decreasing $q$ in a manner consistent with the $q^2$ dependence predicted by OZS. Although our results agree qualitatively with the prediction \cite{OS02} by OZS, the agreement is not quantitative. The reason for the difference between theory and experiment  is not clear at present. On the one hand it might be found in the fact that the calculation is based on a low-order Galerkin approximation. On the other, it could be due to nonlinear interactions between the fluctuations and the use of linearized equations of motion in the theory. 

\begin{figure}
\epsfxsize=2in
\centerline{\epsffile{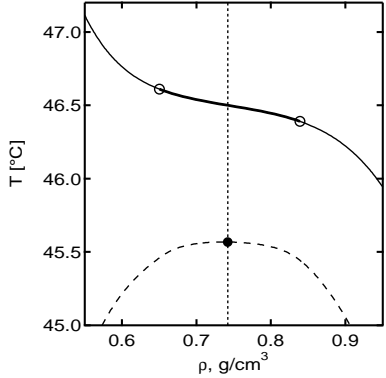}}
\vskip 0.03in
\caption{Temperature-density plane near the critical point of SF$_6$. Dashed line: coexistence curve separating liquid and vapor. Vertical dotted line: critical isochore. Solid circle: critical point $T_c = 45.567^\circ$C, $P_c = 37.545$ bar, $\rho_c =  0.742$ g/cm$^3$. Solid line: isobar $P = 38.325$bars. Heavy solid line ending in the two circles: density range spanned for $\Delta T = 0.220 ^\circ$C.}
\label{fig:phase_dia}
\end{figure}

The apparatus was described in detail elsewhere. \cite{DBMTHCA96}  The top and bottom plates of the cell each consisted of a 0.318 cm thick sapphire plate with a diameter of 10 cm. \cite{FN2} A silver layer of thickness 0.9 $\mu$m was evaporated on the top surface of the bottom plate to provide a mirror for the shadowgraph.  An aluminum plate of equal size containing thermistors and a film heater was pressed from below against the bottom sapphire. The separation $d$ between the sapphire plates was fixed by a porous paper ring with an inner (outer) diameter of 2.5 (3.3) cm. It was measured interferometrically \cite{DBMTHCA96} to be $d = 34.3 \mu$m. Inspection with an expanded laser beam revealed circular Newton rings with their centers close to the center of the paper ring. The actual area sampled by the shadowgraph was near the center of the Newton rings where convection rolls first appeared when $\Delta T$ was increased above the onset at $\Delta T_c$, and had a size of $1.3\times1.3$ mm. Assuming a parabolic profile of the bowing of the top/bottom plates, we estimate that $d$ was uniform to $\pm 0.017 \mu$m or 0.05 \% over the sampled area.
About half the temperature difference across the cell was across the SF$_6$ layer. The measured $\Delta T$ was corrected accordingly. This introduced a systematic uncertainty of $\Delta T$ of about 10\%. Whereas this primarily leads to an overall factor of order unity in the comparison with theory, it makes a sufficiently accurate {\it a priori} adjustment of the mean sample temperature $\bar T$ to the value $\bar T(\rho_c)$ on the critical isochore difficult. Thus we determined $\bar T(\rho_c)$ experimentally.
Figure~\ref{fig:phase_dia} illustrates the phase diagram near the critical point in the $T-\rho$ plane. The pressure was held constant within 0.0005 bar at $38.325$ bars, corresponding to the solid line.  In order to find $\bar T(\rho_c)$, the fluctuation power $P_I$ at constant $\Delta T$ was measured as a function of  $\bar T$ by the shadowgraph method. The value of $\bar T$ at which $P_I$ had a maximum was taken to correspond to $\bar T(\rho_c)$. 
Later, when $\Delta T$ was changed in the experiment, $\bar T$ was held constant at $\bar T(\rho_c)$. Thus the imposed $\Delta T$ caused the sample density to vary vertically between the hot and cold plates as illustrated by the heavy solid line in the figure. 

\narrowtext

The shadowgraph apparatus was a modified version of one described before. \cite{DBMTHCA96} We used a digital camera with 12 bit resolution. \cite{FN3} The light source was a light emitting diode \cite{FN4} behind a nominally $15 \mu$m diameter pinhole. A collimating lens with focal length $f = 5$ cm was used. The shadowgraph transfer function  ${\cal T}(q,z_1)$ was measured by determining the fluctuation intensity at constant $q$ as a function of the optical distance $z_1$  \cite{DBMTHCA96,TC02}. We typically took $N = 1000$ uncorrelated  images $\tilde I_i$ of size $272\times272$ pixels at a given $\Delta T$ and used their average $I_0$ as a reference to compute the signal $I_i = \tilde I_i/I_0 - 1$. The square of the modulus of the Fourier transform $S_i({\bf q}, \Delta T)$ of $I_i$ was used to calculate the structure factor $S_{T,i}({\bf q}, \Delta T)$ of the temperature field of each image. \cite{DBMTHCA96,TC02} The average $S_{T}({\bf q},\Delta T)$ of $S_{T,i}({\bf q},\Delta T)$ over all $N$ images was used to compute the azimuthal integral $S_{T}(q, \Delta T)$. 

For $\Delta T = 0$ the experimentally determined $S_{T}(q, 0)$ was dominated by instrumental noise. Thus we derived the difference $\delta S_T(q, \Delta T) =  S_T(q, \Delta T) - S_T(q, 0)$. It follows from the work of OZS that this quantity is given by 
\begin{equation}
\delta S_T(q, \Delta T) = S^T_{E} \tilde S ^0 _{NE} \tilde S ^R_{NE}(q)\ .
\label{eq:S_T}
\end{equation}
The factor $S^T_{E}$ gives the strength of the temperature fluctuations in the equilibrium system and should not depend on $q$ or $\Delta T$. It contains a number of thermodynamic parameters and Boltzmann's constant. Its estimate from known thermodynamic data is not highly accurate, and we chose to treat it as an adjustable scale factor. According to OZS the strength  of the nonequilibrium enhancement is given by \cite{FN_R}
\begin{equation}
\tilde S ^0 _{NE} = \sigma R + {{(C_P/T)d^2 \Delta T^2} \over {D_T^2}}\ .
\label{eq:S^0}
\end{equation}
Here $C_P$ is the heat capacity at constant pressure and $T$ the absolute temperature. The parameter $\sigma = \nu / D_T$, with $\nu$ equal to the kinematic viscosity and $D_T$ equal to the thermal diffusivity, is the Prandtl number and $R = \alpha_P g d^3 \Delta T / (\nu D_T)$, with $\alpha_P$ equal to the isobaric thermal expansion coefficient and $g$ the acceleration of gravity, is the Rayleigh number. Within the low-order Galerkin approximation used by OZS the dependence of the nonequilibrium enhancement on the horizontal wavenumber $q$ is given by
\begin{equation}
\tilde S ^R _{NE}(q) = (d/36) \Lambda_0^R(q)
\label{eq:S^R}
\end{equation}
with
\begin{equation}
\Lambda_0^R(q) = (30/d) {\cal A} / {\cal B}\ ,
\end{equation}
\begin{equation}
{\cal A} = {{27 q^2} \over { 28 (q^2 + 10)[(q^2+12)^2 + 360] - 27 q^2 R}}\ ,
\label{eq:calA}
\end{equation}
and
\begin{equation}
{\cal B} = \sigma + {{(q^2 + 10)(q^2 + 12)} \over {(q^2 + 12)^2 + 360}}\ \ .
\label{eq:calB}
\end{equation}

\begin{figure}
\epsfxsize=2.1in
\centerline{\epsffile{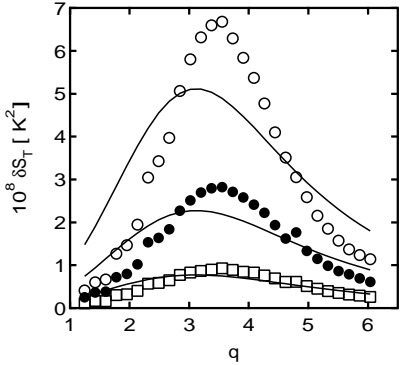}}
\vskip 0.03in
\caption{Experimental results for the structure factor $\delta S_T(q,\Delta T)$ as a function of the wavenumber $q$ for several $\Delta T$. From top to bottom the results are for $\Delta T = $0.189, 0.142, and 0.095 $^\circ$C. For comparison, the onset of convection occurs at $\Delta T_c = 0.440^\circ$C. The lines are fits of Eqs.~\ref{eq:S_T} to \ref{eq:calB} to the data, adjusting only $S_E^T$.}
\label{fig:lin0}
\end{figure}

In Fig.~\ref{fig:lin0} we show experimental results for $\delta S_T(q,\Delta T)$ for several $\Delta T$. Note that all of the $\Delta T$ are well below the value for the onset of convection $\Delta T_c = 0.440^\circ$C. At large $q$ the data show the rapid decrease with increasing $q$ which corresponds to $\delta S_T(q,\Delta T) \sim q^{-4}$ and which had been observed by light scattering. As $q$ decreased, the $q^{-4}$ dependence saturated because fluctuations in the vertical direction were inhibited by the boundaries. At even smaller $q$ the structure factor had a maximum and, with decreasing $q$, vanished. This latter region had not been observed before in experiment, but was predicted by OZS. The maximum of $\delta S_T(q,\Delta T)$ occurred at a value of $q$ which corresponds roughly to the inverse of twice the plate spacing. All the qualitative features are consistent with the theoretical prediction.

At a quantitative level the agreement with theory is not perfect. We fitted the prediction Eq.~\ref{eq:S_T} to the data, adjusting $S_E^T$ separately for each $\Delta T$. The results are given by the solid lines through the data sets in Fig.~\ref{fig:lin0}. Clearly the theoretical curves are much wider than the data suggest. On the other hand, the results for $S_E^T$ were only mildly dependent on $\Delta T$, suggesting that the dependence on $\Delta T$ contained in Eq.~\ref{eq:S^0} is about right.

\begin{figure}
\epsfxsize=2.1in
\centerline{\epsffile{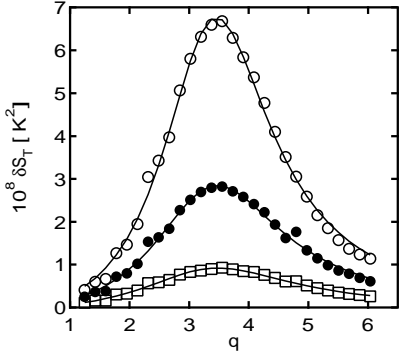}}
\vskip 0.03in
\caption{Experimental results for the structure factor $\delta S_T(q,\Delta T)$ as a function of the wavenumber $q$ for several $\Delta T$. The data are the same as those in Fig.~\ref{fig:lin0}. The lines are fits of Eqs.~\ref{eq:S_T} to \ref{eq:calB} to the data, adjusting the two empirical parameters $f_q$ and $\tilde R$ discussed in the text in addition to $S_E^T$.}
\label{fig:lin}
\end{figure}

\begin{figure}
\epsfxsize=2.1in
\centerline{\epsffile{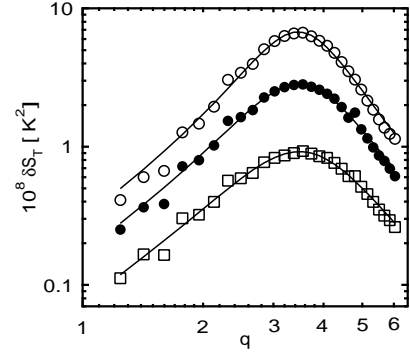}}
\vskip 0.03in
\caption{Experimental results for the structure factor $\delta S_T(q,\Delta T)$ as a function of the wavenumber $q$ as in Fig.~\ref{fig:lin}, but on logarithmic scales.}
\label{fig:log}
\end{figure}

In order to obtain a better fit, we introduced two empirical $\Delta T$-dependent adjustable parameters in addition to the overall scale factor $S_E^T$. First, we allowed an adjustment of the length scale in Eqs.~\ref{eq:calA} and \ref{eq:calB} by replacing $q$ with $f_q q$ and least-squares adjusting $f_q$. This has the potential to expand/contract the curves horizontally along the $q$-axis. Initially this was done to accomodate a possible experimental uncertainty in $d$ which enters into computing $q$. However, this uncertainty allows only values in the range $f_q = 1 \pm 0.03$ and is independent of $\Delta T$. The second adjustment was made in the effective value of $R$ because this parameter influences the widths of the curves, i.e. we replaced $R$ with an adjustable $\tilde R$ in Eqs.~\ref{eq:S^0} and \ref{eq:calA}. \cite{FN_R} We show the resulting fits in Fig.~\ref{fig:lin}. To better reveal the behavior at large and small $q$, the same results are shown on double logarithmic scales in Fig.~\ref{fig:log}. We note that the fits of these empirically modified theoretical predictions are excellent for all three $\Delta T$. From Fig.~\ref{fig:log} we also see that the data agree well with the predicted $q^{-4}$ and $q^2$ dependence at large and small $q$ respectively.

\begin{figure}
\epsfxsize=2.1in
\centerline{\epsffile{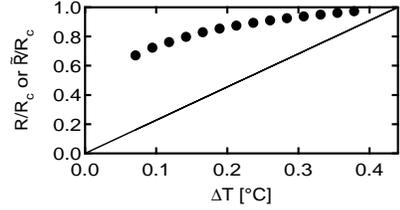}}
\vskip 0.03in
\caption{Experimental results for $R/R_c$ (solid line) and for the fit parameter $\tilde R/R_c$ (solid circles) as a function of $\Delta T$. According to the results of OZS the two should be equal. }
\label{fig:R-tilde}
\end{figure}

\begin{figure}
\epsfxsize=2.1in
\centerline{\epsffile{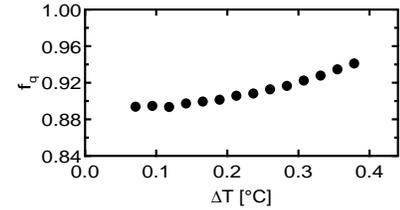}}
\vskip 0.03in
\caption{Experimental results for the length-scale fit parameter $f_q$ as a function of $\Delta T$.  According to the results of OZS $f_q$ should be equal to one.}
\label{fig:f_q}
\end{figure}

\begin{figure}
\epsfxsize=2.1in
\vskip -0.15in
\centerline{\epsffile{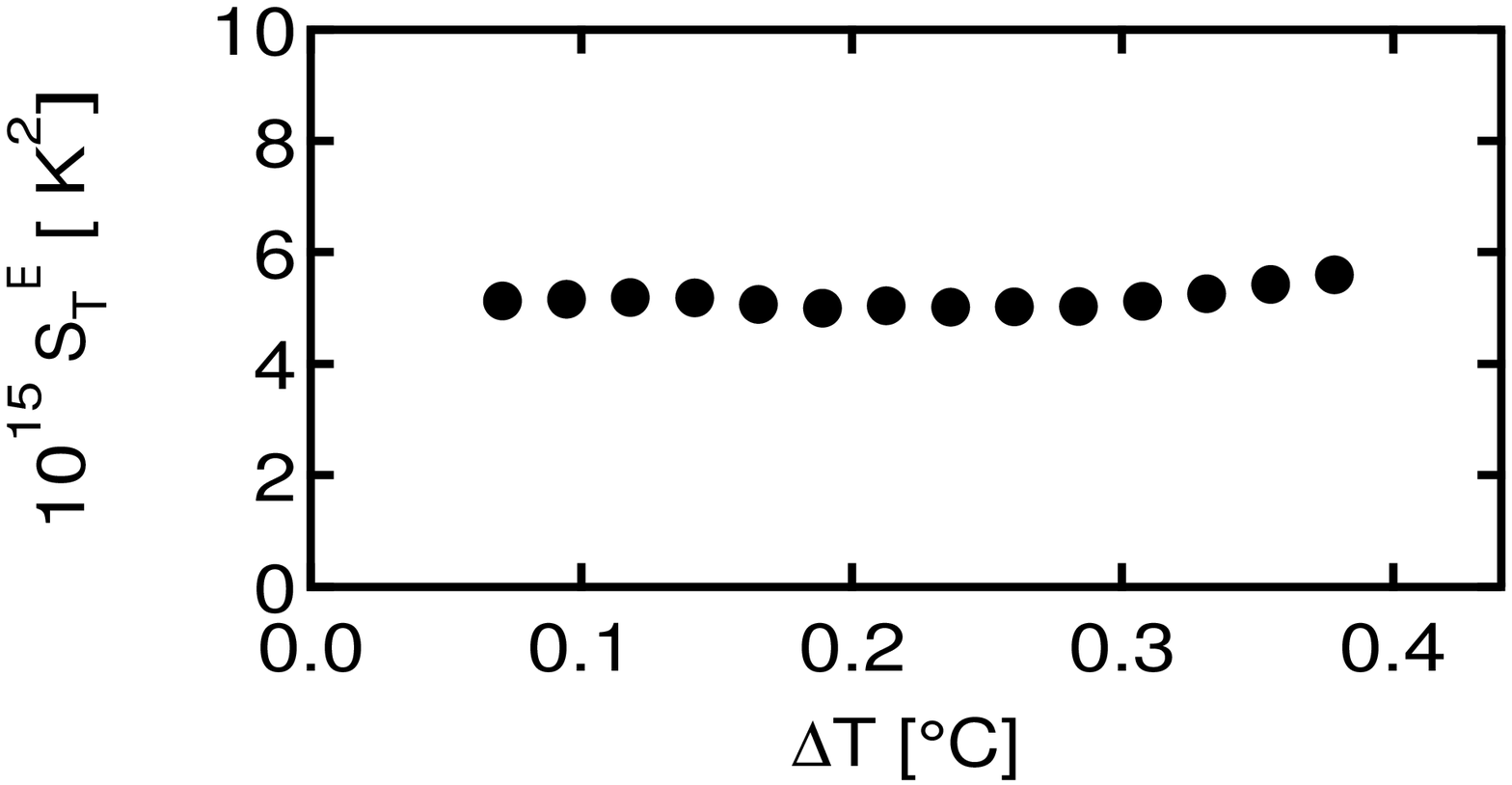}}
\vskip 0.03in
\caption{Experimental results for the coefficient $S_E^T$ as a function of $\Delta T$. According to the prediction of OZS $S_E^T$ should be independent of $\Delta T$.}
\label{fig:SET}
\end{figure}

In Fig.~\ref{fig:R-tilde} we show the results for $\tilde R$ divided by the result for $R_c$ given by the Galerkin calculation of OZS, i.e. by $R_c = 1750$. This parameter primarily influences the width of $\delta S_T(q)$. Also shown, as a solid line, is $R/R_c = \Delta T/\Delta T_c$ based on the experimentally determined $\Delta T_c = 0.440^\circ$C. According to the theory the two should be equal. As $\Delta T$ approaches $\Delta T_c = 0.44^\circ$C, the fitted values of $\tilde R/R_c$ approach the expected value $R/R_c = 1$, but at smaller $\Delta T_c$ there is a large disagreement over a wide  range.

Figure~\ref{fig:f_q} shows the length-scale parameter $f_q$. It is less than the predicted value of unity over the entire range of the experiment. It can reasonably be extrapolated to $f_q = 0.97$ at  $\Delta T = \Delta T_c = 0.440^\circ$C. Within experimental uncertainty this is consistent with $f_q(R_c) = 1$.

Finally, in Fig.~\ref{fig:SET} we present the results for $S_E^T$ which control the overall size of $\delta S_T(q)$. According to the theory this quantity should be independent of $\Delta T$. The experimental results are indeed reasonably constant.

In this Letter we presented experimental results for the non-equilibrium contribution $\delta S_T(q,\Delta T)$ to the structure factor  of a horizontal layer of fluid confined between parallel plates and heated from below. We found that the results have the qualitative features of the predictions of OZS. In particular, at large and small $q$ we found the data to be consistent with  $\delta S_T \sim q^{-4}$ and $\delta S_T \sim q^2$ respectively, with a maximum of $\delta S_T(q)$ at an intermediate  value of $q$ comparable to that given by the inverse plate spacing. To our knowledge the maximum, due to the confinement, and the $q^2$ dependence at small $q$ had not been observed before in
 experiments. A detailed comparison of the shape of $\delta S_T(q)$ reveals quantitative differences between experiment and theory. In order to obtain a fit to the data, we introduced two empirical parameters. One of them, an effective Rayleigh number $\tilde R$, modifies the width of the curves. As shown in Fig.~\ref{fig:R-tilde}, $\tilde R$ differs from $R$ over the entire range of $\Delta T$, with the difference largest at small $\Delta T$. The other, $f_q$, modifies the characteristic length scale. As seen in Fig.~\ref{fig:f_q}, it also differs from the prediction $f_q = 1$  over the entire range, again with the largest difference at small $\Delta T$.

At this point we can only try to speculate about the reason for the difference between theory and experiment. Two possibilities come to mind. On the one hand, the low-order Galerkin approximation which was used may be inadequate. This could be tested by (numerical rather than analytic) higher order calculations. On the other, the neglect of nonlinear terms in the deterministic equations of motion may not be appropriate for the conditions of the experiment. However, in that case we would have expected the disagreement between experiment and theory to diminish as $\Delta T$ decreased. As seen in Figs.~\ref{fig:R-tilde} and \ref{fig:f_q}, just the opposite is the case.

We are grateful to Jose Ortiz de Zar\'ate and to Jan Sengers for stimulating discussions and for providing us with a preprint of their work. This work was supported by the US National Science Foundation through Grant No. DMR00-71328 .

\end{multicols}


\begin{references}
\bibitem{FN2}A diamond-machined Aluminum bottom plate \cite{DBMTHCA96}, with a mirror finish which appeared perfect to the eye, caused the fluctuations to align with the tool marks.
\bibitem{FN3}QImaging Model Retiga 1300. For the measurements reported here we used an exposure time of 0.2 sec. Measurements with 0.5 sec yielded smaller values of $S_T(q)$ and thus of $S_T^E$, but the shapes of $S_T(q)$ (i.e. the results for $\tilde R/R_c$ and $f_q$) were not influenced significantly.
\bibitem{FN_R} The contribution $\sigma R$ in Eq.~\ref{eq:S^0} is negligible for our cases.
\end{references}
\end{document}